# Configuration Work: Four Consequences of LLMs-*in-use*

**Gabriel Alcaras**

médialab, Sciences Po

Paris, France

gabriel.alcaras@sciencespo.fr

https://orcid.org/0009-0008-9267-7559

**Donato Ricci**

médialab, Sciences Po

Paris, France

donato.ricci@sciencespo.fr

https://orcid.org/0000-0002-4739-1622

## Abstract

This article examines what it means to use Large Language Models in everyday work. Drawing on a seven-month longitudinal qualitative study, we argue that LLMs do not straightforwardly automate or augment tasks. We propose the concept of configuration work to describe the labor through which workers make a generic system usable for a specific professional task. Configuration work materializes in four intertwined consequences. First, workers must discretize their activity, breaking it into units that the system can process. Second, operating the system generates cluttering, as prompting, evaluating, and correcting responses add scattered layers of work that get in the way of existing routines. Third, users gradually attune their practices and expectations to the machine's generic rigidity, making sense of the system's limits and finding space for it within their practices. Fourth, as LLMs absorb repetitive tasks, they desaturate the texture of work, shifting activity toward logistical manipulation of outputs and away from forms of engagement that sustain a sense of accomplishment. Taken together, these consequences suggest that LLMs reshape work through the individualized labor required to configure a universal, task-agnostic system within situated professional ecologies.

## Keywords

Generative AI; LLMs; work; use; configuration; consequence; impact

# Introduction

It's quite remarkable that, for all the discourse surrounding Large Language Models (LLMs), almost none of it concerns the ordinary ways they are woven into everyday work. In fact, we seem so uninterested in their use that we would rather speculate about their future impacts on the job market, comment on the latest anecdote, such as careless lawyers who submitted ChatGPT-fabricated cases[1], or argue about what should or should not be done with them, than attend to what people actually do with LLMs at work. ChatGPT is often heralded as the "fastest-growing consumer internet app of all time"[2] and surveys suggest that 40 percent of the US population uses generative AI (Bick et al., 2024), but we still know little about what it means to "use" this technology, and what, if anything, it changes in work practices.

Much of this conversation is driven instead by renewed automation anxiety, fueled by claims that "this time, it's—really—different" (Wajcman, 2017). In a widely cited paper, OpenAI, the company behind ChatGPT, claims that 46% of jobs "could have over half their tasks affected by LLMs" (Eloundou et al., 2024). Anthropic, a competitor to OpenAI, estimates that around 36% of occupations use "AI for at least a quarter of their associated tasks" (Handa et al., 2025). These studies can certainly measure the extent to which people use LLMs, assess their distribution across various tasks, and distinguish between automation and augmentation. They also overlook how LLMs are integrated into daily routines, shaped by specific practices and the needs and constraints of different professional environments. Ignoring these factors obscures how workers actually use and experience these technologies, as well as the consequences for them and their professional practices.

This study contributes to contemporary research on artificial intelligence and work (Deranty and Corbin 2024; Lei and Kim 2024) by examining the consequences of LLMs for work practices, drawing on lessons from the ethnography of algorithmic systems (Christin 2020) and perspectives from Science and Technology Studies (STS). We rely on a

---

longitudinal qualitative study organized around four cohorts, each comprising eight workers engaged in a structured, seven-month participatory protocol. Since LLMs are predicted to have the most significant impact for younger, entry-level workers (Brynjolfsson, Chandar, and Chen 2025), we present findings from our first cohort (September 2024 to March 2025), consisting of eight final-year Master's students actively involved in professional environments, through internships, consulting projects, or research collaborations, with the ongoing analysis of subsequent cohorts providing converging observations. We base the analyses presented here on 30 hours of group meetings, 8 hours of individual interviews, 400 pages of participants' written notes, and 600 LLM conversations.

We demonstrate that LLMs do not seamlessly integrate into work practices. People must instead *make them work* through the practical labor required to turn a generic system into one usable in specific professional ecologies. We call these efforts configuration work. The four consequences we develop extend this notion by attending to the often-unattended practices and experiences through which workers configure both the machine and their own activity. In doing so, we aim to expand the collective analytical repertoire for describing what it means to use LLMs.

## Impact or Consequences: How Do We Account for AI in Work?

The current wave of automation anxiety surrounding AI is only the latest episode in a much older history under industrial capitalism (Mokyr et al., 2015). The revival of the "end of work" narrative is emblematic: despite substantial evidence that its predictions are overstated (Deranty & Corbin, 2024; Lei & Kim, 2024; Shestakofsky, 2017, pp. 378–83), it continues to animate the public debate, helped along by AI companies. Their willingness to openly acknowledge—and sometimes advertise—the potential harm of their products is, in fact, a distinctive feature of contemporary AI controversies, setting them apart from earlier ones (Marres et al., 2024). This strategic acknowledgement enables these companies to appropriate and control critical discourses (Dandurand et al., 2023), positioning them advantageously in regulatory discussions while reinforcing public perception that their technologies are efficient, even dangerous. Impact studies respond directly to this atmosphere. By quantifying how AI might displace, augment, or reorganise work, they aim to inform debate and guide policy, and, in doing so, have been prominent in public discourse.

Current studies on LLMs in professional settings focus on assessing their impacts—the measurable changes resulting from their implementation—such as how these computational technologies reshape productivity, employment, or worker welfare. Impact estimation seeks to capture the causal relationship between technological adoption and three specific phenomena: displacement, when machines replace human labour; augmentation, when AI automates subtasks or enhances human performance, enabling workers to increase productivity; and renewal, when new forms of work and skill emerge in response (Acemoglu & Restrepo, 2022).

Most impact studies are grounded in the "task model of labour markets" (Autor et al., 2003). This framework conceives jobs as bundles of tasks: distinct, observable activities that together define an occupation. The methods used to determine the automation susceptibility of a task are, of course, critical—and highly variable. From surveys in which workers assess the automatability of their tasks (Shao et al., 2025), to computational methods that infer task exposure from prompt–task relations (Chatterji et al., 2025; Eloundou et al., 2024; Handa et al., 2025), and even to macroeconomic analyses (Acemoglu et al., 2024), these studies rely on a shared foundation: structured task ontologies such as the Occupational Information Network (O*NET), which catalogues roughly 19,000 tasks and 1,000 occupations. Each task is then examined to determine whether it is subject to displacement or augmentation, based on relative efficiency, cost, and technological capability. If a substantial share of an occupation's tasks is judged susceptible to substitution or augmentation, the occupation is classified as highly exposed. The cumulative effect of these task-level assessments across all occupations constitutes the labour-market impact of technology.

The task model has brought consistency, comparability, and "greater descriptive realism" (Acemoglu et al., 2024, p. 2) to impact studies, marking a clear improvement over far less reliable approaches that produced headline-grabbing estimates (Frey & Osborne, 2017). However, its own advocates have highlighted important challenges, from the difficulties of task-based labour-market estimation to the risks of forecasting future technologies based on current AI progress (Autor, 2013). From our perspective, the task model's limits also come from what it leaves out of view. By treating work as the sum of discrete, stable tasks, it cannot account for the routines that involve extended periods, the distribution among multiple actors and technical systems (Hutchins, 1995), the informal and tacit dimensions of work (Polanyi, 1967), the dynamic interplay between different tasks and their articulation within organizations (Strauss, 1988), and how activities play into the broader ecologies of professions and their jurisdictions (Abbott, 1988). Likewise, the emphasis on AI capabilities encourages a deterministic reading of technological change, one that extrapolates from what systems could do while leaving aside how they are actually enacted in work practices.

As a result, impact studies remain oriented toward measurable outcomes (productivity gains, automation risks) while offering little insight into how these technologies reshape the dynamic processes and experiences through which work is actually done. As is often the case in economics, the task model risks being performative, thereby making these abstractions a reality and a dominant perspective in public policy, debate, and research (Fourcade et al., 2015). An alternative approach should attend to the ways workers make sense of computational technologies within their own ecologies of practice.

We therefore call for an examination of the consequences of LLMs on work. Where impact studies measure large-scale outcomes, model causal mechanisms, and look into the future, consequences describe situated processes, attend to the mutual shaping of workers and technologies, and remain grounded in the present. Following pragmatist thought (Debaise, 2005; Stengers, 2005), we define consequences as ongoing negotiations: on the one hand, an unfamiliar technology enters into new relations with an existing situation and gradually reconfigures what comes next; on the other hand, the situation answers back to the arrival of this new entity and configures its place within establishes routines, norms and ecologies. To describe consequences is, then, to observe how these negotiations take shape in everyday practices, how workers experience them, and how they open or foreclose possibilities within these situations.

The lack of consequence studies regarding AI, and more specifically LLMs, stands in stark contrast not only to its widespread use, as noted earlier, but also to the number of impact studies on the topic. We know very little about the concrete ways knowledge workers and experts engage with these systems (Lei & Kim, 2024). Studies of AI systems more broadly, though not focused on LLMs, have examined how automation reshapes specific professional domains, with radiology standing as the most developed case (Lombi & Rossero, 2024). Others document the rise of algorithmic management (Kellogg et al., 2020), the effects of cheaper, but not necessarily better, automated systems on workers' pay and conditions (Vieira, 2020), and the ways AI unsettles implicit norms and generates controversy within organisations (van den Broek et al., 2020). Qualitative research on LLMs themselves remains thin and relies primarily on qualitative interviews or surveys, focusing on the expectations of early-career professionals (Vicsek et al., 2025) or early-adopter practices (Retkowsky et al., 2024).

To contribute to this nascent field of consequence studies, we examine LLMs-*in-use*. This perspective purposefully moves away from the hegemonic perspective of innovation in favor of what matters to users (Edgerton, 1999). Instead of starting from issues attributed to the systems' internal functioning (e.g. bias, explainability, or confabulation), we observe what users construe as problems and how they do so, what other unforeseen challenges or creative uses might arise in specific occupational settings. Consequently, studying technology *-in-use* (Suchman et al., 1999) and *-in-practice* (Orlikowski, 2000) and requires ethnographic accounts of enactment within specific situations (Knorr Cetina, 2014) in which meaning, power, and agency continually emerge through everyday adaptation and response between humans and machines.

## Designing a Studio to Describe the Consequences of LLMs-*in-use*

The framing of technology-*in-use*, however valuable for the study of LLMs in work practices, presents numerous difficulties for practical, qualitative inquiry. First, LLMs exacerbate the individualization of digital services, which ensures that no two users encounter the same feed or interface, and the progressive closure of public access, as platforms restrict APIs, curtail scraping, or make data retrieval impracticable (Tromble, 2021). Generative AI sits at the acme of these two trends, since they deliver individualized outputs that respond to users' prompts, while operating behind the closed doors of proprietary services.

This problem of access is compounded by issues of social acceptability. A notable feature of GenAI adoption, often obscured by its rapid pace, is its bottom-up character: workers introduce these systems into their own routines of their own volition. Much of today's use in the workplace is therefore unsanctioned and happens without managerial oversight—organizations often frame this as "shadow AI." As a result, many workers either conceal their use or downplay it when asked, especially when organizational guidelines discourage certain GenAI practices. The broader discourse surrounding LLMs further saturates their use with normative expectations. In some professional settings, relying on an LLM can be labelled as unethical, lazy, or seen as incompetence. As several of our participants noted, this produces a climate of self-censorship in the workplace, with workers underreporting their practices and withholding those deemed more problematic.

A third challenge concerns the increasing instability of digital technologies. Contemporary computational tools are introduced into everyday life long before they stabilize. Since the early 2000s, venture capital imperatives have encouraged a permanent state of beta (Neff & Stark, 2004) in a constant pursuit of innovation (Vinsel & Russell, 2020). Tests are no longer confined to bounded settings but intervene directly in the social world (Marres & Stark, 2020). This instability is heightened in the case of LLMs, whose recent and still-uncertain adoption multiplies hesitations about what they are and what they do. For researchers, this complicates the inquiry, as the object of study keeps evolving.

To address these challenges, we propose designing a studio. Just as the factory offers a vantage point on industrial labour and the laboratory on scientific practice, the studio serves as our vantage point on how LLMs unfold in ordinary work. There, workers participate in our study as co-inquirers (Gourlet et al., 2024) to foreground the precise situations in which LLM use arises (Figure 1). To address fragmented and closed access, co-inquirers bring conversation logs, work experiences, and recollections into the studio, contributing both digital traces and the contextual knowledge needed to interpret them. To accommodate bottom-up adoption and limit self-censorship, the studio provides a setting outside the workplace where co-inquirers can redeploy their practices, express hesitations, and speak openly. Finally, to address technological instability, the studio functions not as a one-off session but as a recurring meeting space, allowing co-inquirers to track their use over extended periods.

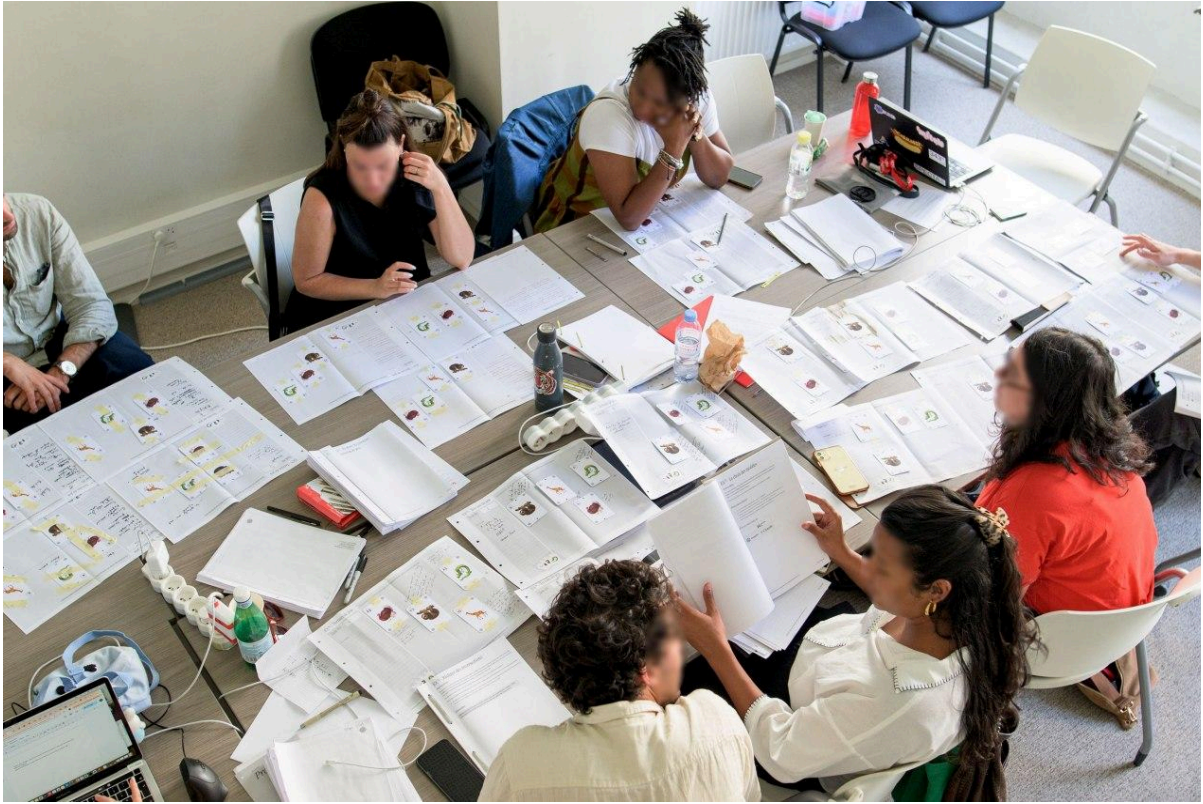

**Figure 1.** Meeting in the studio (June 18, 2025). Photo by the authors, EL2MP project.

Inside the studio, co-inquirers follow a research protocol conceived as a form of ethnographic invention and intervention (Criado & Estalella, 2023; Farías, 2025), in which work practices are deliberately tensed (Ricci, 2019) to make the role of generative AI easier to discern. The protocol begins with individual semi-directive interviews to document participants' working conditions, practices, and overall stance toward LLMs. It then guides participants through a series of exercises to revisit their use, dissect their routines, and reflect on their practices (Ricci, 2022). These exercises range, for example, from reviewing and annotating conversation histories, including forms of self-quantification, to comparing outputs generated by identical prompts across different models, to graphically representing their experience when pursuing a specific goal. We open-sourced and published the full protocol as a 120-page printed vademecum (Alcaras et al., 2025). This research device augments descriptions, granting access to aspects of practice that would otherwise remain inaccessible or under-specified to an external observer. At each return to the studio, co-inquirers brought the results of one or several exercises (Figures 2 and 3). These written manuscripts functioned both as data and as supports for discussion, providing reflective surfaces against which participants could articulate what happens in LLMs-*in-use*. This alternation between asynchronous individual work and collective analysis structures the protocol, which typically lasts six to eight months.

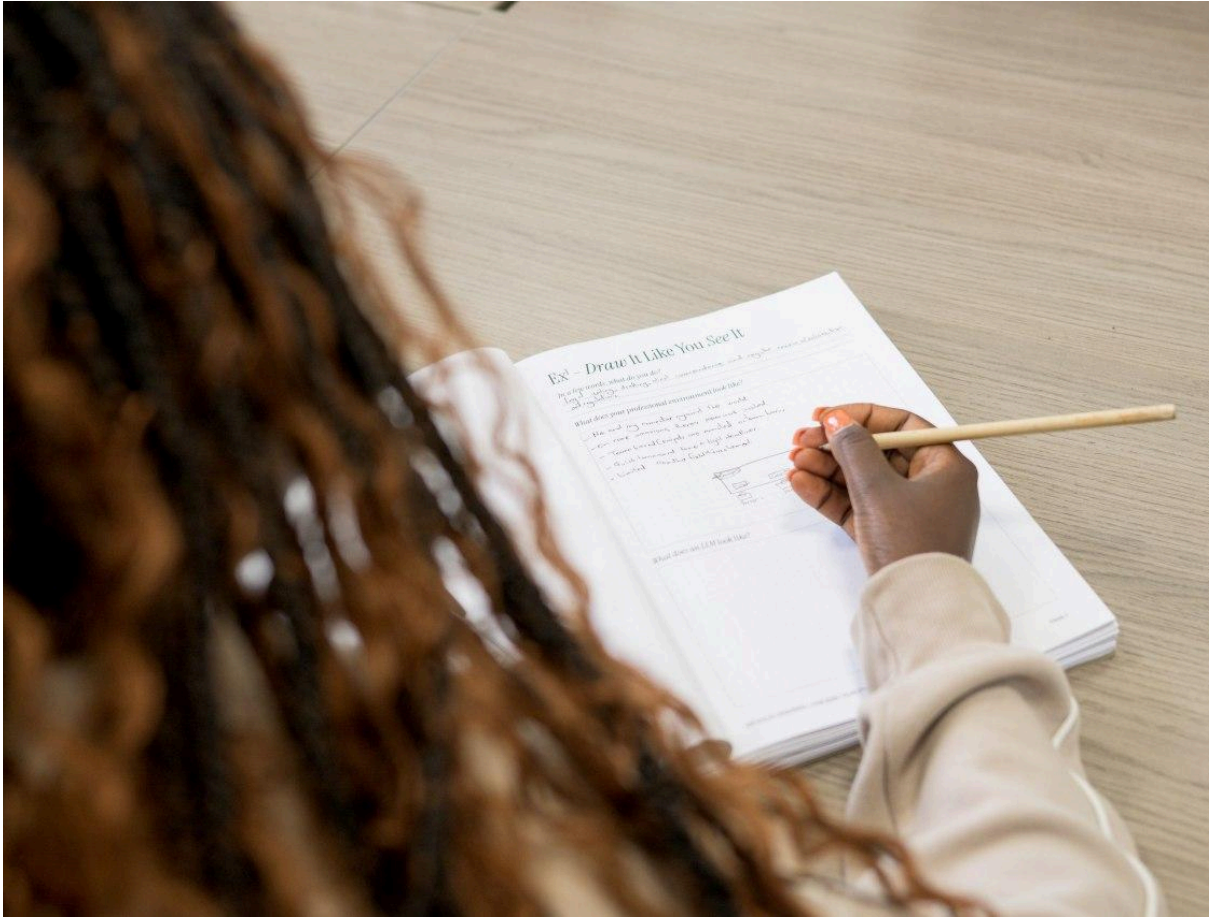

**Figure 2.** Participant working with the vademecum during a studio session (October 9, 2025). Photo by the authors, EL2MP project.

We implemented this protocol across four cohorts, each composed of eight participants, for a total of 32 co-inquirers. Participants met with us regularly in the studio throughout their cohort. We adopted a cohort-based design for practical reasons and to foster within-group discussion, in which contrasting experiences encouraged reflexivity and thus richer descriptions. We alternated between cohorts of Master's students and cohorts of working professionals, with Cohort 1 (September 2024–April 2025) and Cohort 3 (September 2025–April 2026, ongoing at the time of writing) comprising students, and Cohort 2 (January–October 2025) and Cohort 4 (January–October 2026, due to begin shortly at the time of writing) comprising, respectively, mid-career artists and cultural producers and mid-career civil servants. This article presents findings from a systematic analysis of the first cohort's output, corroborated by material from cohorts 2 and 3.

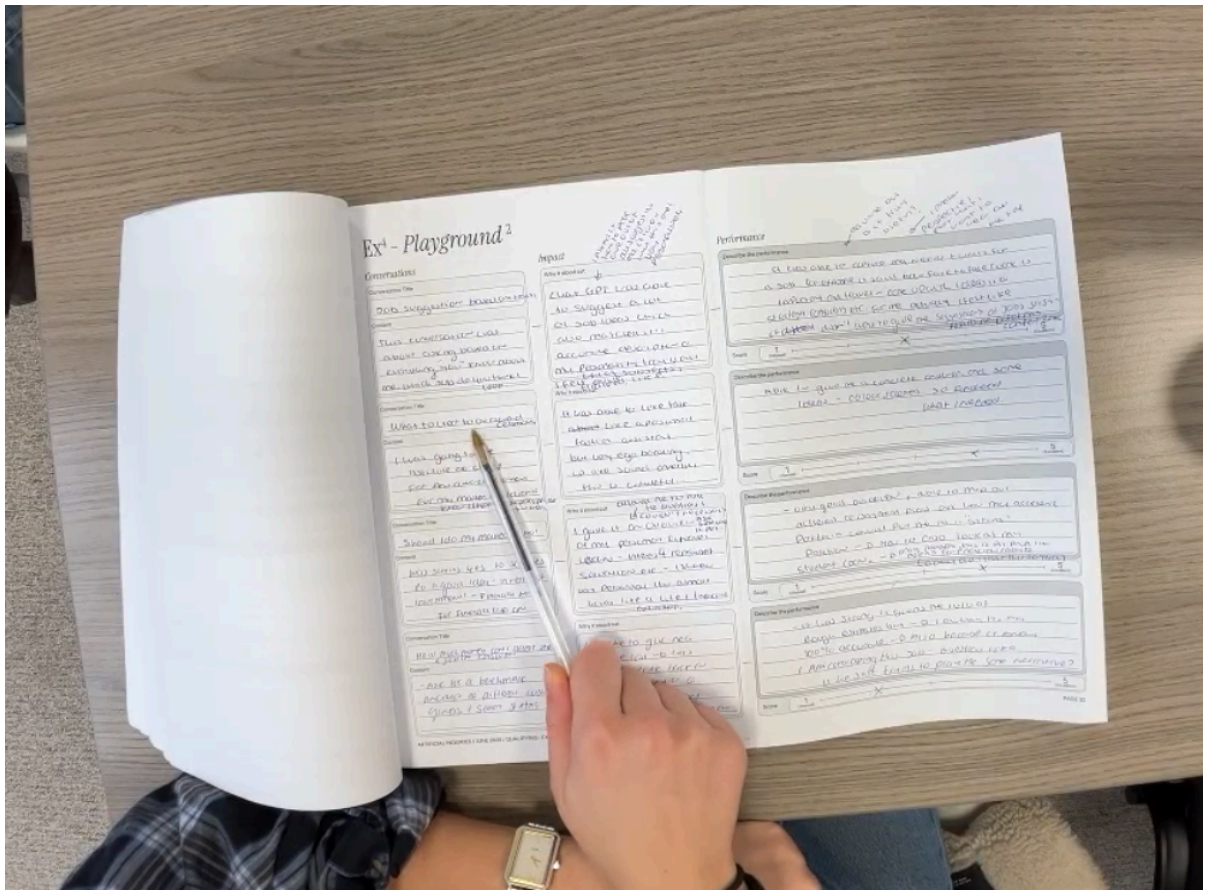

**Figure 3.** Screenshot from a top-down video recording of a studio session, showing annotated pages of the vademecum in use. Photo by the authors, EL2MP project.

The cohort comprises eight final-year Master's students who are actively engaged in professional environments through internships, consulting projects, or research collaborations (see Table 1). Located in the heart of Paris, Sciences Po is a French elite institution (Brown et al., 2016). It is one of the few higher-education institutions in France that charges tuition and is known for producing some of the country's top politicians, journalists, and entrepreneurs. Graduates typically enter high-paying professional roles in knowledge-based sectors and, on average, earn twice the salary of their peers in the same age group. We received dozens of applications and carefully selected participants to form a diverse group that could meet regularly. All students in the cohort were compensated to ensure adherence. We provided each participant with a paid professional account for the most widely used LLMs (ChatGPT, Claude, Mistral) to ensure greater data privacy. As the group included two German students, one Chinese student, and five French students, we used English for group discussions.

**Table 1.** First cohort overview: field of study and prior work experience

| Participant | Field | Professional Experience |
| --- | --- | --- |
| Alice | Law | Legal intern in business and art law |
| Camille | Law | Legal intern in litigation and corporate criminal law |
| Yichen | Public Policy | Research assistant for global tech regulation |
| Tobias | Public Policy | Project assistant for development programs |
| Charlotte | International Affairs | Internships in international agencies focused on migration |
| Agnès | International Affairs | Internship at an embassy |
| Constance | Economics | Research assistant; production intern for a radio program |
| Guillaume | Urban Planning | Intern in territorial development for renewable energy projects |

Once the cohort completed the protocol, we proceeded inductively from the material, drawing on grounded theory (Glaser & Strauss, 1967) and collective interpretation in the spirit of data sessions (Joyce et al., 2023). Our corpus included 30 hours of audio-video recordings of studio meetings, 400 scanned pages of the vademecum, more than 600 LLM conversations donated by participants, and 8 hours of individual interviews. As a research group, we first convened in collective data sessions in which we reviewed and annotated the transcripts and recordings, beginning with the interim and final reflections—presentations in which participants drew on concrete examples to articulate how LLMs had intervened in their routines. Building on these insights, we returned to the remaining materials (vademecum pages, donated conversations, interviews) to test what had emerged collectively and to surface elements that had been absent from the group discussion.

We identified four consequences, all linked to what we call configuration work, which we develop in the next section. Drawing on traditions of ethnographic vignettes (Bloom-Christen & Grunow, 2024) and praxiology (Mol, 2002), we present each consequence as a dialogue between two texts: an ethnographic scene, offering our account of participants' experiences, discussions, and lines of thought; and our commentary, where we progressively articulate and refine our concepts by moving back and forth between the scene and relevant strands of the literature. We also organized a re-reading session with the cohort, circulating draft texts for annotation before convening a collective discussion. This closed loop allowed participants to contest or extend our interpretations, making the write-up itself another moment of joint inquiry.

## Configuration Work and the Consequences of LLMs for Work Practices

We argue that using an LLM entails a specific form of practical labor: the effort required to turn a generic system into one that is usable within a particular professional ecology. We call this labor configuration work. As is typical with the tech industry, AI companies oversell the capacities of their technological solutions and conceal the involvement of human workers (Irani, 2015; Shestakofsky, 2017). Indeed, without configuration work, conversational LLMs linger as general-purpose artefacts; they rely on users to bring them into the situation, to shape their orientation, and to interpret whatever they produce. Their task-agnostic character (Schulz-Schaeffer, 2025) shifts the burden of specification onto users, who must construct the conditions under which the system becomes relevant. In this sense, LLMs challenge a classic STS tenet: the scripted artifact (Akrich, 1992). Configuration work suggests that, while they might not be scripted in the canonical sense, conversational LLMs operate in practice as if they scripted their own re-scripting by users, who must repeatedly re-specify the machine. Just as articulation work underpins the negotiated order of organizational life (Strauss, 1988), configuration work underpins the negotiated use of LLMs within individual practice. In what follows, we explain how configuration work manifests itself through four consequences.

## Discretization

The first consequence for work is what we call discretization: breaking down work into units that the LLM can process and respond to. Despite the promise that these systems can do *anything*, workers must still instruct them to do *something*. Discretization arises from the material conditions of interaction, shaped by the constraints of the interface, the technical limits of the system, and the norms of professional work.

---

**Scene 1. Discretization**

Most participants turned to LLMs for tasks that were clearly defined. Interestingly, they had already been using other services—Google and Wikipedia for research, DeepL for translation, Stack Overflow and documentation for help with coding. ChatGPT replaced them all, akin to a convenience store. Participants appreciated the time it saved and the comfort of staying within a one-size-fits-all platform.

Participants also built on the existing division of labor in their field to assign tasks to the LLM. Researchers quickly sought to test the accuracy of literature reviews or bibliography management. Yichen found ChatGPT "incredibly efficient, especially in the writing of [quantitative] social science essays since they tend to be standard in format." At the same time, Charlotte noticed that "they're quite good when it comes to referencing with rigid guidelines. So if there's something really clear, like a formula for how to do referencing, for example, it usually can do that." Most agreed that LLMs performed best on standardized tasks, governed by precise rules.

But not all uses were simple substitutions. Some participants relied on what LLMs were "good at," depending on what they had picked up from the media, friends, or colleagues. Nearly everyone used ChatGPT to "paraphrase." Using it for coding purposes was also widespread, justified by the idea that "it's good at it" because they had heard it had been referencing on a lot of code. Yet for them, coding was mostly seen as a means to an end. It didn't need to be elegant; it just needed to be functional. Constance said she didn't care if her code was pretty or fast—only that it ran. Tobias went further, rejecting the idea of coding as "creative" work:

"I think for me, it's more like implementing something I've already set out. [...]. It feels like I've already put the thought into how I wanted to do it, and then I just tell it [ChatGPT] to implement."

Over time, however, participants began to push beyond these obvious use cases. In law, for instance, Alice realized that what she thought of as a single task was too unwieldy for the LLM. What felt like one smooth action to her had to be broken into smaller, more manageable steps:

"I think one of the key things to get proper results is really to dissect every part of what you want to reflect on and go step by step. Because even if you give the big picture at the beginning and then try to break it down, it tends to try to do everything at once. I mean, it was really confused."

---

Camille, a lawyer, viewed many of her tasks as long-term processes, cumulative in nature, which didn't translate well to the kind of short-term, local, iterative problems LLMs seemed able to tackle:

"LLMs have difficulty following a progressive and logical approach over the long term. At first, I thought 'OK, law mostly relies on logic [...] so LLMs could be really good at it'. But working with them showed me that it's often very hard for the machine to build on previous versions, to integrate feedback. And since my work required that kind of continuity, sometimes the long process became the most frustrating aspect of the project. [...] That's a more global kind of reasoning. [...] And I think the LLM doesn't really do that, because it responds to specific prompts and tasks. So when, at the end, you ask for a global answer, it struggles."

Although this was often disappointing, it also reminded participants that their jobs weren't so easy to automate. "I thought my work would be simple," Guillaume reflected. "Something so standardized that the LLM would handle it quickly. But actually, it wasn't." The following week, Camille echoed him: "I think it shows that law, even when it looks like a simple legal task, always contains hidden complexity."

Listening to Camille, Charlotte introduced a metaphor that resonated with many: working with an LLM felt like trying to fit your job into a small window. Even the act of framing the task could feel like a struggle:

"Yeah, I think... Do you ever feel that too? Because I keep struggling with it, like... The window to put in the information and all the context just feels so small. Like, sometimes, you don't even know how to fit everything in. The struggle already starts with just trying to frame it."

Fitting information within the window was challenging because it wasn't just a matter of providing more context to the LLM; too much could backfire. Constance noticed that overloading the prompt "creates noise in the response—like it's farther from what I actually want than if I just provide one document with a more precise question." This insight led her to change strategy: "I didn't change the way I prompt, but I did change how I use the chats. Now I use way more separate chats to get more precise responses."

LLM use builds on the prior discretization of work. Many of the tasks delegated to LLMs are well-defined activities that were already outsourced to other services or embedded in existing divisions of labor. New tools seldom produce the dramatic transformations their inventors claim; instead, they adapt to existing structures and practices. Use can act as a "suppression of radical change": cinema, for example, strengthened the theater's position as a dominant site of entertainment (Winston, 1996); the telephone, in its early years, deepened existing social patterns rather than creating new ones (Fischer, 1992). Workplace studies point to the same dynamic. The introduction of CT scanners in hospitals revealed that organizational consequences depend less on the technology itself than on its integration into institutional histories (Barley, 1986). New tools are often enacted through "limited use", resulting in the "reinforcement and preservation of the structural status quo, with no discernible changes in work practice" (Orlikowski, 2000, p. 421).

What, then, is distinctive about the form of discretization we observe with LLMs? The association between technology and the division of labor is hardly new, dating back to the Babbage Principle (Braverman, 1998) to the current taskification of work highlighted by platform studies (Poutanen et al., 2019). In this scene, we observe three features that distinguish the discretization associated with LLMs. First, it operates at a different level. Fragmentation of work is typically described as occurring between workers (Friedmann, 1964), either through horizontal or vertical divisions of labor, or across different temporalities. With LLMs, by contrast, fragmentation occurs within the individual's own activity. Second, the sequence of discretization is reversed. In earlier cases, decomposition was a prerequisite for automation, built into the organization of labor or into machines themselves. With LLMs, discretization can happen *after the fact*, in use, as work is reframed to fit within the constraints of the interface and context window. Third, the burden of discretization shifts. Rather than being imposed externally by managers or engineers, the responsibility falls to the worker, who must decide how to parse their own work for the system.

Discretization thus refers to the individualized, after-the-fact fragmentation of work, where the responsibility for making activity legible to the system lies with the worker. It is not an abstract principle but a practical effort, which may explain why most participants preferred pre-discretized tasks that already fit the system and spared them additional work. Interestingly, because discretization requires noticeable effort, participants became aware of the many aspects of their work that could not be made to fit easily within LLMs. Participants sometimes turned this into a new standard of measure: work was judged in terms of whether it could be discretized or not. In this sense, discretization could also be a process of valuation, prompting workers to recognize the intricacies of activities that had become so routine as to be mundane. Such realizations are not confined to expert work; similar dynamics have been observed on the assembly line, where apparently simple operations reveal layers of hidden skill and coordination (Pfeiffer, 2016). The attempt to automate these routines often makes their significance visible: for example, efforts to automate classroom roll-call prompted teachers to defend it not as a record-keeping task, but a valued moment of relational work with students (Selwyn, 2022).

## Cluttering

The second consequence, which we call cluttering, refers to the accumulation of additional forms of work to make the machine do the job, sometimes taking space away from actually doing it. Whereas discretization concerns existing work, cluttering addresses the effects of the machine introducing new, small practices that feel out of place, unnecessary, or hard to navigate.

---

**Scene 2. Cluttering**

No one enjoyed using LLMs. This caught us off guard, as we had expected at least some participants to have fun coaxing better answers from the machine. Guillaume was the ideal candidate who could enjoy the process. He had gained a bit of a reputation in the group for how playfully and inventively he pushed the machine's limits. But even he declared:

"There's no joy in it, because I don't want to prompt the machine to do it. I just want the machine to do it. I don't want to have a role in the process. My ideal AI would be the one that automatically knows when to send an email, sends it, and just gets it out of my head."

Prompting was a chore, something to skip whenever possible. The goal was to paste content, press Enter, copy the output, and move on. Reaching that point still required some setup: a brief prompt at the start of the conversation, tailored to a task such as coding, translation, or paraphrasing. However, participants rarely bothered with careful prompting unless something went wrong. On the topic of getting help with the LaTeX language, Constance confided: "I didn't try to perfect my prompts for this type of task, as the responses were generally satisfactory—even with very implicit formulations." A review of their prompts confirmed this carelessness. Participants acknowledged that they contained numerous typos, were hastily written, and were awkwardly phrased.

Some tasks inevitably required more carefully worded prompts. This usually occurred when participants deliberately tested the LLM's limits or believed that initial effort would yield general prompt templates that could be reused in other conversations. Agnès discovered early on that crafting precise prompts tends to produce better results, making the initial effort a worthwhile investment: "I tend to do quite detailed prompts because I want the LLM to be effective. I really put a lot of information in it. When [...] I asked more general questions, I got a lot of hallucinations with ChatGPT."

Evaluating the LLM's output posed another challenge. As participants encountered more frequent confabulations, they grew increasingly wary of their answers. While the machine sped up the act of writing, it multiplied the work of reading. With chatty models like ChatGPT, the vigilance required became exhausting. Many participants were concerned about the sustainability of providing such constant attention, fearing that fatigue or ambiguity might erode their capacity over time. Evaluation became even trickier when the LLM convincingly mimicked professional voices. Camille noted ChatGPT's ability to sound lawyerly, and Constance felt it could pass as a typical economist, making errors harder to catch.

Participants coped by turning towards tasks that were easier to evaluate. Some, like Charlotte, preferred tasks with "a certain baseline of reference", in which they had enough prior experience to judge the result quickly. For others, tasks outside their expertise offered relief, either because they didn't care or knew less about them. Unburdened by expert knowledge, their evaluation used more straightforward, cruder criteria, making it less cognitively demanding. Coding was one such task: they deemed it mechanical, non-creative, and easy to test, as code either runs or doesn't.

However, professionals do not evaluate outputs solely based on their expertise. They're often simultaneously trying to understand how the output relates to their input, so that they could craft better prompts, get better results consistently and thus minimize the burden of evaluation. Unfortunately, many considered the machine inconsistent in its responses. Agnès describes her frustration as a "feeling of absurdity": "[In the beginning], I thought I could kind of adjust how it worked. But in the end, I found that the more I tried to get a precise formulation, the more random the results became. [...]. I had no way of knowing how to influence the outcome." Constance felt the same way: "I didn't have the tools to understand the breakdown, and so I couldn't solve the problem." They concluded they couldn't understand the machine well enough, or that it was fundamentally unreliable in its behaviour. Either way, many stopped caring about crafting better prompts. At best, they relied on simple, ritual phrases they believed would yield the best possible results with little effort. Tobias explained:

"While it is difficult to determine whether it actually improves the outputs, [...] I especially love [the] emotional stimuli [technique], because no matter the prompt, just adding 'this is very important for my work' already leads me to believe I put sufficient effort in maxing out the LLM."

After a certain point, prompting and evaluation consumed too much time and space relative to what they provided. Camille captured this common experience:

"There's this moment when I realize, after giving multiple instructions, clarifying, or rephrasing, that ChatGPT is giving me completely off-topic information. I end up feeling really frustrated and give up, out of lack of time and motivation. [...] ChatGPT just doesn't understand what I ask, despite all my efforts."

Making the machine work clutters daily routines with many new practices (writing prompts, assessing outputs) that participants perceive as cumbersome. This diffuse layer of labor feels out of place in part because it contradicts users' expectations. While the role of expectations has been widely examined in shaping science and technology (Borup et al., 2006), research also shows that users' expectations contribute to the imagining and enacting of affordances (Nagy & Neff, 2015). Here, expectations provide a frame of reference against which the necessary effort becomes visible and judged as misplaced. LLMs promise that they would free professionals from gruesome tasks and allow them to spend more time on what they consider essential, rewarding work. So when they clutter work routines, these activities feel "dirty" (Hughes, 1984): necessary, yet falling outside what workers recognize as the proper boundaries of their occupation. Interestingly, just as delegation is one way to manage dirty work, cluttering gives rise to a range of decluttering tactics—choosing tasks that require minimal prompting or evaluation, relying on short ritual prompts, or using mechanized forms of verification such as running code to see if it executes.

Cluttering also hints at the sense of disorder that LLMs introduce into everyday work. Indeed, participants felt that the system's responses were not closely aligned with their prompts and were somewhat arbitrary. This unpredictability echoes *automation surprises* (Sarter et al., 1997): moments when automated systems behave in ways that defy user control or expectation. In this scene, surprise is not fleeting novelty that diminishes with familiarity, but a sustained mode of experience in which the machine's behavior continues to catch users off guard even after repeated use. As a result, participants could not turn the play of input and output into what they would call an expertise. This contrasts with more predictable forms of automation in fields such as radiology, whose future-oriented professional identity leads practitioners to view AI as increasing the complexity of their work and, thus, their expertise (Lombi & Rossero, 2024). With LLMs, automation surprises make users feel like unproductive labor, where effort does not yield a proportional transformation of the written output, making it difficult for workers to experience use as skilled or satisfying.

What makes clutter distinctive is the impossibility of achieving a sufficient understanding of the system—and therefore control—through repeated use. The metaphor is no longer the black box (Ziewitz, 2016), but the black hole: it is possible to invest significant effort in the LLM and get little in return. In response, users adopt decluttering tactics. While some operate within the system (such as using short, ritual prompts that sustain the belief that they are maximizing the machine's potential), many others involve boundary work, centered on deciding when to engage with the system, when to stop prompting, and when to withdraw from it altogether. This form of self-management recalls the individualization observed in discretization. Because participants felt that the unpredictability of LLMs prevented their use from becoming a site of skill or mastery, these systems fostered neither excellence nor sloppiness, but rather *good enoughing* (Bialski, 2024): a practice through which actors determine what counts as satisficing rather than optimal use.

## Attunement

The third consequence for work manifests as attunement: the gradual adjustment of practices, expectations, and valuation to the machine's perceived generic rigidity. If cluttering refers to the scattered work of using the system and managing its shortcomings, attunement, by contrast, describes how users come to situate it within their existing ways of working and professional ecologies.

---

**Scene 3. Attunement**

As months pass, participants grow more confident that ChatGPT won't replace them. Some even begin to wonder if LLMs were ever genuine contenders. Initially, many treated the AI as a fellow expert. Now, they speak to it as a novice or an alien—someone unfamiliar with their field. Camille realized this the day she noticed how similar prompting felt to explaining the basics to a clueless and stubborn client, rather than collaborating with a seasoned colleague:

"The most difficult part of legal reasoning is reformulating. When you have a client coming in with a question that's all over the place, and you have to figure out what the actual problem is, and then explain it. And with the LLM, it was kind of the same, because we always had to explain it again. You can't really assume that you're talking to a lawyer."

The core challenge lay in choosing the right words, whether it was articulating what felt off in the machine's reply or describing precisely what they wanted. In fact, participants often felt at a loss for words. Even a driven user like Guillaume arrived at a point of wordlessness: "I just didn't really know what to say to it anymore." Agnes described a similar dead-end when she realized she didn't know enough about the subject matter to address the LLM adequately: "when the result is too general, and I don't know enough about the subject to ask more precise questions, I feel like I'm at a dead end, because I can't choose a new path of questions." In fact, participants preferred quick roleplay-based prompts —"you're a social media marketing expert"— rather than a detailed spelling out of what constituted expertise. It was easier "to make GPT-4 believe it's an expert on the topic" (Tobias) than doing the work of context-setting, which "takes a lot of energy and thought I am sometimes too lazy to do" (Charlotte).

However, their main frustration isn't that the AI falls short of professional standards. It doesn't seem able to adapt. No matter how carefully they prompt or how much context they provide, the LLM fails to improve or behaves unpredictably. During a group discussion, Camille compared LLMs to interns. Sure, they can handle repetitive tasks, like sorting and renaming files, but they don't "know" how to behave. Unlike interns, LLMs don't learn through context or observation. Camille puts it bluntly: an intern would "revise her emails five times" without being told. An LLM has to be reminded of instructions incessantly.

---

Others nod around the table. Agnes recalled her own internship at an embassy. When Charlotte challenged her—"Colleagues make mistakes too, so how are they different from ChatGPT?"—Agnes replied:

"I think I distrust the machine more, and maybe I'm just biased, because it's probabilistic—I really don't think it understands. But if it's a colleague or an intern, that person can still learn, and you can actually teach them how to do it."

Many agreed: it's not the mistakes that bother them, it's the lack of learning. This disconnect limits the machine's capabilities. Yet, for some, its very distance from their professional world makes it valuable. Alice, for instance, appreciates that ChatGPT isn't part of her social or work circle. That separation creates a space that feels private and free of judgment. As she once told the group:

"It's kind of a tool you can use anytime, day or night. So you develop a certain kind of interpersonal relationship with the LLM, and it feels safe to ask it any question, even the kind of question you might feel stupid asking someone else. You don't feel like you're going to be judged afterward, even if you say something dumb."

This type of task was so prevalent and important to Charlotte that she named it "silly work." In a moment of vulnerability, she confided to the group that to calm her anxiety before phone calls, especially in English, her second language, she would ask ChatGPT for a short script. She rarely used it, but just having it there made her feel prepared. The ritual itself mattered more than the result. She explained:

"I always feel a bit weird telling people that I still write myself notes before making a phone call. It's like, yeah, maybe in the professional world there's this kind of judgment. Like, you can't even make a call without prepping? So that's why it feels lower stakes to do it with something like ChatGPT than to just do it on my own. I could do it myself, but it would take a lot of time. And I'd probably feel a bit guilty spending so much time on a task that, in the end, maybe I don't even need, because I often don't even look at the notes that much. But because it gives me a sense of security, and because it's fast with ChatGPT, it kind of resolves that tension."

Though Charlotte was the first to label it, we had encountered this kind of "silly work" long before in others' private written logs. Constance and Agnes also described turning to ChatGPT for reassurance, precisely because it wasn't part of their social world. For Constance, an economist who felt intimidated by programming, the LLM offered patient support, helping her regain her confidence. For Agnes, the value wasn't in doing more or faster, but in feeling at ease: you do not go "further" with the LLM, just "more serenely."

Attunement is itself a consequence of the machine's generic rigidity. Users experience this rigidity through discretization (making their work fit) and cluttering (working for the machine), progressively working around the LLM constraints. Studying technology-*in-use* offers a different perspective on a widely noted feature of generative AI: its task-agnosticity. Whereas previous designed technologies absorbed labor capacities through a codified capture of knowledge for an imagined user and an imagined task, machine learning introduces another form of absorption, called emergence: systems generalize from data, and thus lack task-relatedness, ushering in a new "universality of the machine" (Steinhoff, 2024). Yet the same genericity that made the LLM an everything machine also prevented it from fitting within the particularities of real work practices. Participants found that it often lacked the specificity required to integrate into their professional ecologies, even when they attempted to correct and adjust the LLM—task-agnostic rather than ecology-specific.

Attunement manifests that users recognize that the LLM is not infinitely malleable, and this unresponsiveness is interpreted as a distinct machinic behaviour. Just as social scientists have noted the presence of machinic bias through computational exploration of these models (Boelaert et al., 2025), participants reach a similar conclusion based on their own experience. Indeed, many drew on analogies from their professional worlds to make sense of the machine's role, comparing the LLM to an intern and themselves to its manager, thereby extending a long lineage of managerial class and ideology (Chandler, 1977). While many felt that the analogy was apt, it also highlighted that, unlike an intern, who can be socialized, an LLM requires that every norm, context, and expectation be made explicit. This dependence on explicit instruction through language reveals how much of professional knowledge is tacit, embodied, or distributed (Collins, 2012). This scene exposes the difficulty of locating what feels off or uncanny in a given output—and, even more, of finding words to correct it. While it might not be impossible in principle, such explicitation often demands extraordinary effort, and rarely becomes routine practice.

In the end, attunement is the participants' effort to find a workable stance toward the LLM. The process resembles what AI developers call *alignment*, yet it operates in the opposite direction. Alignment seeks to tune the machine to human values; attunement captures how workers reconfigure their practices in response to the machine's limits, learning to live and work with it. Over time, participants came to create a space for the LLM where the machine could be relied on without being fully trusted, useful without being knowledgeable to their ecologies. In this sense, attunement refers to what participants value when using the LLM. Beyond the instrumental value of producing more or faster, the LLM provided participants with a sense of reassuring presence, a suitable companion for low-stakes, affective, or "silly" work.

## Desaturation

Desaturation names the fading of color in work. As LLMs absorb the most formulaic and repeatable tasks, a change happens in the texture of labor, as it loses differentiation and alters the distribution of agency. Workers find themselves managing outputs rather than crafting them, doing more of what is easiest to automate and less of what feels engaging, leaving a flat surface of logistical operations—copying, pasting, and circulating fragments of text.

---

**Scene 4. Desaturation**

For our final session, we asked our eight participants to reflect on their six months of using LLMs. What emerged was a story of ambivalence. Many spoke of disenchantment, almost literally so: "The LLM is not magic," said both Guillaume and Tobias. Others hoped to offload tedious work and save time. But half a year later, they admitted that the LLM didn't live up to that promise. And yet, most had come to "rely" on it for many things and confessed they'd feel its absence. This admission made some uneasy, as reliance raised uncomfortable questions about their skills and relationship to their work.

That discomfort arose from a growing sense that LLMs were, in some way, reshaping their work. The machine had become a container for the "boring," "uninteresting," "uncreative," and "unenjoyable." Gradually, it absorbed every task they didn't like. Lumped together in ChatGPT, the least fulfilling parts of their job became more visible. As Charlotte explained:

"I think throughout all the practice, I really found myself gravitating toward some boring tasks—like we all did. Things I don't really enjoy, and that often take up way too much time, things I don't actually want to spend time on. For me, that's usually paraphrasing."

Some didn't feel freed by this. The LLM made tedious tasks easier, so they found themselves doing more of them. Instead of clearing time for "creative" work, they sank deeper into what the tool could handle. Agnes, a researcher, noticed this: "I'm spending too much time on the [literature] review part [...] and I should stop that, and focus more on the thinking part." Constance also worried that LLMs would divert economists from the work they *should* be doing:

"A large part of our time as economists is spent on really uninteresting tasks. That's probably why economists have so many research assistants [laughter]. And also, the profession is relying more and more on heavy tools that require tons of annotation, data cleaning, and all that. Which means less time for more theoretical or analytical tasks. So, in theory, LLMs could help gain time for that. But in practice, I don't think that's what's happening. Because now that we can do more complex things with machines, we end up pushing the tools even further. [...] We just want a

---

fancy method that will impress people. [...] So it's true, it expands possibilities a lot for economists. But is that really what we should be doing? I don't know."

LLMs didn't just change what participants did, but also how they experienced it. As the discussion went on, Guillaume expressed that using the LLM for all these tasks made them bland:

"Before [using the LLM], I used to do those tasks already, I didn't have more or less work. It just wasn't boring. I mean, I didn't think of it as something super interesting, but it wasn't boring either. And with the use of the LLM, it started to feel more and more boring. [...] It's not just that it revealed something [about my work], but somehow, the way we use it creates the boredom."

Constance has reached a point at which she systematically relies on the LLM for all her coding tasks. While she appreciated the help, she acknowledged that achieving "results" is not the same as feeling "accomplished". Instead of being the source of her work, she became a conduit, an "interface":

"And I think that's why I have some issues with the work I'm producing, because I feel like I'm just an interface for code that's already been written. [...] So I'm happy when I get the final results. But in the meantime, during those long days of doing it, I don't really feel very accomplished, I'd say. Whereas I remember, when I was starting to code without ChatGPT, every time I managed to do something, it felt like a real event."

Some were concerned that this transformation had occurred gradually, insidiously. It crept in unnoticed and, using the LLM, became second nature. As Tobias put it: "I don't know if there's really that much emotion involved. Honestly, it's more just a matter of habit." Months before this discussion, Constance had written in her log: "The machine is now part of my daily life, whether I use it or not." Even unused, the LLM loomed in the background. Deciding when to use it became part of the work itself. Some came up with discursive strategies to justify their use. Tobias, for instance, drew a personal line between tasks that required agency and those that didn't. Others framed it as a matter of self-discipline, like Agnes:

"I have a very short deadline, like a month and a half, and I'm super, super late on my work. [...] I'm trying not to let the time pressure get to me. I tell myself: no, I'll take the time I need. [I'm trying to set boundaries] for which tasks I do myself and which ones I might do with ChatGPT. But it's a slippery slope. At the beginning, I didn't want to use it at all. But then I got stuck [...] so I asked GPT. [...] And I'm trying not to. But because it works, it's hard not to. Still, I don't want to be productive just for the sake of productivity. I just want to do good work."

Desaturation concerns the *how* of work. It's one of modern "ironies of automation" (Bainbridge, 1983): the LLM automates rote work, yet makes the work itself feel more boring. This stems in part from the uniformity of the conversational interface. Tasks once performed through distinct gestures and affordances now converge into a single repetitive loop of input and output. With fewer differentiated actions, the experience of work loses its texture, and workers lose a sense of control over their activity. The LLM's generic rigidity erodes the small handholds through which users can intervene, control, or sense progress. Participants rarely described LLM use in terms that evoke "true work" (Bidet, 2011), the process that transforms both the object and the subject through doing. They felt their practices drifting from productive to logistical (copying, pasting, transferring, monitoring text) where they no longer left traces of themselves in the fruit of their labor. The loss of texture is a process of estrangement, a form of subjective alienation, that even being creative, excellent, and masterful in prompting could not avoid. This scene suggests that configuration work is not only about a surplus of labor, but also about "the process through which subject/object relations are reworked" (Suchman, 2006, p. 262), altering the distribution of active and passive roles in the interaction between worker and machine.

Desaturation also addresses the consequences of the LLM on *what* gets done. It addresses transformations in the ecology of practice and the concerns of practitioners who experience them. Through discretization, decluttering and attunement, participants felt that the machine redirected work towards what could be automated with ease, what was "good enough" – which often was, as we discussed, the type of work they valued or cared about the least. Desaturation emerged among participants who felt that certain practices within their professional ecologies were already under significant strain. They feared that LLMs would compound these pressures, put a finger on the scale, and further endanger practices they held dear or considered important. What also concerned participants was the unintentional nature of this redistribution of work. The use of the LLM had become habitual, in the sense of smoothing out the discontinuities of working practices, the hesitations, the conscious decision-making. Automation had led to automatism, a practice without attention and intention.

## Conclusion

What does it mean to "use" an LLM at work? We argue that it consists in configuration work, through which workers make a generic system usable in a particular professional situation. The question then becomes: what is being configured in configuration work?

The four consequences we identify demonstrate that configuration extends beyond merely prompting the machine. Through discretization, the task itself becomes something to configure. Workers are encouraged to break parts of their activity into units that the system can take up, drawing either on existing discretization or on fresh cuts intentionally made to fit the machine's narrow aperture. Cluttering adds another dimension to configuration work by showing how taxing it can become; the LLM sometimes resembles a black hole that absorbs attention and effort. It forces a recalibration of adequacy, as workers decide how much configuration is worth undertaking to achieve results that will suffice. Attunement offers yet another reworking, this time of the roles assigned to human and machine. Faced with generic rigidity, workers readjust how they speak to the system, what they assume it can do, and how they position themselves in relation to its limits, seeking a workable stance rather than mastery. Desaturation, by contrast, refers to the alteration of the ecology of practice. As workers increasingly manage outputs rather than craft them, the activity drifts toward what is easiest for the system to automate, increases logistics over production, and dulls the texture of everyday activity. Workers do not just configure the machine but what counts as human and machinic, redistributing agency, attention, and the balance of active and passive work along the way.

As noted earlier, our aim in this study is to contribute to an analytical repertoire that enriches and sharpens descriptions of LLMs-*in-use*. The concepts we advance are not meant as mechanistic or obligatory components of using LLMs, nor do we claim that every worker will encounter all of them, or in the same way, or that these exhaust the range of possible consequences. They are the categories that most effectively captured the practice and experience of LLM use in the first cohort of our research protocol. In subsequent cohorts, we aim to examine further consequences, including what workers value in these tools. Over time, we will also need to understand how these consequences vary across occupations, education levels, gender, and other social variables. For now, given the paucity of studies examining LLMs-*in-use*, we believe this repertoire offers a valuable starting point, as without an account of configuration work, the gap between idealized use and situated practice will remain out of reach for analysis.

Throughout the paper, we have shown that a technology heralded as revolutionary partly reflects familiar histories of automation and existing professional routines. This close examination also reveals what might be distinctive about LLMs. A first salient transformation is not automation or augmentation *per se*, but the individualization that comes with configuration work. Traditionally, fragmentation and scripting are understood as deliberate, pre-configured operations, organized through managerial control or embedded in technical design. LLMs seem to introduce a new condition where these processes happen *after the fact* and become the worker's individual responsability.

Second, for some workers, unpredictability and inconsistency prevent LLM use from becoming a site of expertise or satisfaction at work.

In contrast to more deterministic technologies, where familiarity and training tend to crystallize into skill, participants repeatedly encounter automation surprises. They cannot transform experience into dependable control or a satisfying feeling of responsiveness. Third, the value of the LLM is not reducible to the quality of its results. Many participants did not seek perfect outcomes; they relied on outputs that were merely good enough to advance their work. The accuracy of the outputs tended to matter even less outside workers' professional jurisdiction, either because they lacked the knowledge to evaluate it, judged the stakes to be low, or regarded the task as strictly executional, a simple translation of their prompts. Ultimately, the value of LLMs was partly instrumental, but it extended beyond doing more, faster. Participants also valued the LLM as a reassuring presence, one that often felt less fraught to approach than a colleague. These outcomes reveal as much about LLMs as about the working conditions and professional hierarchies they encounter, reflecting the contemporary workplace.

The research presented in this article is part of the Ecologies of LLM Practices (EL2MP) project. With support from Google.org. Further information is available at https://ecologiesofllm.medialab.sciencespo.fr.

## Acknowledgements

This research would not have been possible without the work of Tommaso Prinetti and Zoé de Vries as part of the Ecologies of LLM Practices (EL2MP) project. Special thanks to the organizers and participants of the Data & Society workshop "What Is Work Worth?" (May 8–9, 2025) for their careful and stimulating discussion of an earlier draft of this paper, as well as our colleagues Jean-Samuel Beuscart, Sylvain Parasie and Veronica Barassi. We are also grateful to the participants in the "Méthodes inventives, design participatif, collaborations expérimentales et gâteaux artisanaux" workshops for their sustained engagement in the protocol and data sessions. Finally, we owe a particular debt to all the co-inquirers for their time and generosity, especially the first cohort.